\documentstyle[preprint,eqsecnum,aps]{revtex}
\begin{document}
\draft
\def \beq{\begin{equation}}
\def \eeq{\end{equation}}
\def \beqarr{\begin{eqnarray}}
\def \eeqarr{\end{eqnarray}}

\title{Meron Pseudospin Solutions in Quantum Hall Systems}
\author{Sankalpa Ghosh and R. Rajaraman\cite{byline1}}

\address{School of Physical Sciences \\
Jawaharlal Nehru University\\ New Delhi 110067, \ INDIA\\ }

\date{\today}
\maketitle
\begin{abstract}

In this paper we report calculations of some  pseudospin
textures for  bilayer quantum hall systems with filling factor $
\nu =1$. The textures we study are  isolated single meron
solutions.  Meron solutions have already been studied at great
length by others by minimising the microcopic Hamiltonian
between microscopic trial wavefunctions. Our approach is
somewhat different.  We calculate them by numerically solving
the nonlinear integro -differential equations arising from
extremisation of  the effective action for pseudospin textures.
Our results can be viewed as augmenting earlier results and
providing a basis for comparison.Our differential equation
approach  also allows us to dilineate the impact of different
physical effects like the pseudospin stiffness and the
capacitance energy on the meron solution.

\end{abstract}

\section{Introduction}

Recent years have witnessed the demonstration of topologically nontrivial 
spin textures like Skyrmions as physically important 
excitations in quantum Hall systems. It has been argued in a body
of very interesting papers that such Skyrmions are not mere
exotic theoretical curiosities but will actually be the lowest
energy excitations in some situations ( such as for example filling factor
$\nu = 1$ and small Zeeman coupling)
\cite{Sondhi}. Furthermore,  experimental evidence indicating the
presence of such Skyrmionic excitations has also subsequently emerged \cite
{Barrett}. Skyrmions of  topological charge more than unity have also
been studied \cite{Sondhi2}. 

These results were basically for mono-layers
of electron gas in strong magnetic fields, but with spin degrees of freedom
of the electrons treated dynamically. Meanwhile,
the remarkable discoveries of the quantum Hall effect,
originally found in single two dimensional electron layers, have
also been extended to bilayer systems , thanks to the
development of techniques for growing GaAs heterostructures
containing two separated layers of two-dimensional electron gas
(see for example references \cite{Eisen}) .  On the theoretical front,
a large body of work has already been done on bilayer systems.
An extensive list of references to this literature has been
given in the lucid review of this subject by Girvin and
MacDonald \cite{GirvMac} and in the paper by Moon et al
\cite{Moon}. 

To simplify matters, one may  begin analysing bilayer systems by
assuming , to start with, that the electrons are spinless, i.e.
that the spins of the electrons are fully polarised by the
strong magnetic field and frozen as degrees of freedom.  The
idea would be to later incorporate the complications brought
about by  spin degrees of freedom  after the spinless problem is
understood. However, even when real physical spin is suppressed,
it is helpful to view electrons in a bilayer system  as carrying
a "pseudospin"  in addition to their spatial coordinates on the
plane. The notion of such a pseudospin arises from mapping  the
bilayer spinless problem into a monolayer problem with spin
\cite{Mac}.  In this mapping the layer degree of freedom is
specified by  a two-component pseudospinor whose up and down
components refer to the amplitude for the electron to be in the
first and second layers respectively. Such a mapping allows one
to borrow for  bilayer systems, the rich body of insights and
results available from single layer systems with real spin.  For
instance, just as in case of electrons with real spin  the
combination of the Pauli principle and Coulomb interactions can
lead to a ferromagnetic ground state, one can similarly expect
the pseudospin of bilinear systems to also be fully "polarised"
under corresponding circumstances. Such psuedospin magnetisation
amounts to spontaneous phase coherence between the two layers,
with many interesting consequences , discussed in the literature
\cite{GirvMac},\cite{Moon}.

Going beyond the ground state, the relevance of Skyrmions to
 systems with real spin has in turn
prompted studies of  similar topological excitations in spinless
bilayer systems, but now involving \underbar{ pseudospin} (See
references \cite{GirvMac},\cite{Moon}and references given
therein).  Because of interplane-intraplane anisotropy in
pseudospin stiffness  in bilayer systems, as well the
capacitance energy of carrying unequal charge density in the two
layers, the effective Action governing  pseudospin enjoys only
U(1) symmetry of rotations about the z-axis (the direction
perpendicular to the x-y plane  of the layers).  Finiteness of
energy requires that asymptotically the pseudospin must lie on
the easy (x-y) plane.  The basic topological excitations in that
case are the so-called merons which are vortices in pseudospin 
with a winding number of one-half, and  meron anti-meron bound
pairs.  Such a pair is topologically equivalent to Skyrmions
and carries unit winding number. (For an introduction to such
topological excitations, their winding numbers,  requirements 
due to finite energy,  etc. see reference \cite{Raj}.)

	In this paper we obtain  single meron solutions in
bilayer systems with filling fraction $\nu=1$, for a range of
different values of system parameters (such as the inter layer
separation $d$) by numerically solving the integro differential
equation obeyed by them. Even though a single meron by itself
has a logarithmically divergent energy in the infinite volume
limit it is nevertheless an object of interest.  Physical meron
anti-meron solutions which do have finite energy, are often
treated as a bound pair of two extended interacting objects,
namely the individual merons each with a "size" and a meron core
energy .  Merons are not rigid bodies but rather solutions of
nonlinear wave equations; the notion of giving each of them a
fixed size, shape and energy, although only approximate when
they appear as a part of a larger solution containing  other
merons, is nevertheless useful in such Soliton physics.   Single
meron solutions allow us to obtain their size, shape and
individual energy in an unambiguous way.
 
  Single meron and bi-meron solutions  have already been
extensively studied  in a body of papers by Girvin, MacDonald
and co-workers  \cite{Brey} \cite{Yang} and \cite{Moon} . These
calculations are based on optimising microscopic wavefunctions
with respect to the microscopic interaction Hamiltonian. The
emphasis is  on determining the optimised energy of bimeron
configurations and extracting individual meron core sizes and
energies from them.  Our work here relies heavily on the
advances already made in these papers, and is to be viewed as
something which  will hopefully augment their calculations.  We
will obtain explicit meron solutions through a somewhat {\it
different} route , by numerically integrating the nonlinear
differential equations derived from the effective action obeyed
by the pseudospin magnetic moments. Meron profiles are of some
interest in their own right. [It is for instance not widely
realised that unlike Skyrmions which exist even in the pure
nonlinear sigma model (NLSM) limit, that model will not permit
exact meron solutions. Meron solutions require departure from
the NLSM model, which is provided by the capacitance energy and
other anisotropies. We will explain this point further later
on.] We will obtain our meron profiles for a range of values of
the layer distance and study the way in which the solution is
affected by the addition of leading corrections to the NLSM
effective action. We also compare our results for the meron size
and core energy with those obtained from other methods.
 
\section{Recapitulation of Theoretical Preliminaries}

The integro-differential equation  which we will integrate to get
our meron solutions is obtained by extremising an effective action
which has already been derived by Moon {\it et al}  \cite{Moon}
 starting from the 
basic  microscopic physics. See also Ezawa \cite{z}. For completeness , 
and to set up the notation let us quickly recall their result .
 The pseudospin texture of a state is
described by a classical {\it unit} vector ${\vec m}(\vec r )$ which gives the 
local direction of the pseudospin. Here  $\vec r $
is the coordinate on the x-y plane carrying the layers, while the magnetic
 field B is along the z-direction. The fully
polarised "ferromagnetic" ground state $| \Psi_0 \rangle $ corresponds 
to ${\vec m}$ pointing everywhere in the same direction, say,
 along the x-axis. Using this as the reference state, any other
  state with some arbitrary  texture ${\vec m}(\vec r )$
  is given by performing a local pseudospin rotation on this uniform
  ground state :
 \beq |{\vec m}(\vec r ) \rangle \ \equiv \  \ e^{-i \cal{O}} \
  |\Psi_0 \rangle \eeq 
Here the pseudospin rotation operator is, for infinitesimal rotations,
\beq {\cal O} \ = \ \int d^2 x  \ \big({\hat x} \times  \ {\vec m}(\vec r )
\big) \  . \ {\vec S} ({\vec r}) \eeq
where ${\vec S}$ is the local pseudospin operator given by $ {\vec S} =  \ 
{1 \over 2} \psi^{\dag}_{\alpha} \ \sigma_{\alpha \beta} \ \psi_{\beta}  $
 and $\psi_{\alpha}$ is the second quantised electron field operator
with $ \alpha$ denoting the pseudospin (layer) index.
The effective action $I[{\vec m}]$ for any given pseudospin texture
$ {\vec m}(\vec r )$ has already been
derived by Moon {\it et al}  \cite{Moon} starting from the 
basic  microscopic physics. See also Ezawa \cite{z}. The microscopic
 Hamiltonian is

\beq H \  = \ H_K \ + \ H_{int} \eeq
where
\beq H_K \ = \ \frac{1}{2m} \sum_{\alpha=1}^{2}\int d^{2}x  \ 
\psi^{\dag}_{\alpha} \ D^{2} \ \psi_{\alpha} \label{HK} \eeq
 and,
 \beq H_{int} \ = \ \frac{1}{2}\int d^{2}x 
d^{2}x' \ \delta \rho_{\alpha}({\bf x}) \ V_{\alpha\beta}(\bf x-x') \ 
\delta \rho_{\beta}({\bf x'}) \label{HI}\eeq
and $\delta \rho_{\alpha}$ is the  local density operator minus the
 mean density in the $\alpha$th layer.The interaction potential
  written here as a $2 \times 2$ matrix in 
pseudospin(layer) space is due to
 Coulomb forces between electrons in
the same or different layers and is given by
\beq V_{\alpha\beta}(\bf x-x') \ = \ 
\frac{e^2}{ \epsilon \sqrt{ ({\bf x - x'})^{2} + 
d^{2}(1 - \delta_{\alpha\beta})}}  \label{int} \eeq
where $d$is the layer separation.

The effective action $I[{\vec m}]$ is just the expectation value of 
the microscopic Hamiltonian in the state $ |{\vec m}({\vec r }) \rangle$,
 but with all operators projected on to the  lowest landau level (LLL).
The detailed calculation is given in ref \cite{Moon} and leads to a nonlocal
action functional of $ {\vec m}({\vec r })$. Upon doing a gradient expansion
, valid for slowly varying (long wavelength) textures, the leading terms
are
\beq
I ({\vec m})=\int d^{2}r  \ \bigg[\frac{1}{2} \rho_{A} \big(\nabla
m_{z})^{2} + \frac{1}{2} \rho_{E} \big((\nabla m_{x})^{2} + 
(\nabla m_{y})^{2}\big) + 
  \beta m_{z}^{2} \bigg] \ + \ C[{\bf m}] 
\label {Eff} \eeq
where \beq C[{\bf m}] \ \equiv {e^{2}d^{2} \over 32\pi^{2}\epsilon}
\int d^{2}r\int
d^{2}r'({m_{z}({\bf r})\nabla^{2}m_{z}({\bf r'}) \over |({\bf
r}-{\bf r'}|})
\eeq
and terms of order $ \nabla^3 $ and higher have been dropped.
The constants appearing in this effective action are given in terms 
of microscopic interaction parameters. The constants $\rho_A$ and 
$\rho_E$ are pseudospin stiffness parameters whose physical origin is the
exculsion principle (Hund's rule) mentioned earlier. They are given by
\beqarr \rho_A \ &=& \ \big( {\nu \over 32 \pi^2}\big) \int_{0}^{\infty} dk 
 k^3 \  V^A_k \ exp({-k^2 \over 2})    \nonumber \\
        \rho_E \ &=& \ \big( {\nu \over 32 \pi^2}\big) \int_{0}^{\infty} dk 
 k^3 \  V^E_k \ exp({-k^2 \over 2})   \label{rho}  \eeqarr
where $V^A_k \ = \ 2\pi e^2 /(\epsilon k)$ and  $V^E_k \ = \  exp(-kd) 
(2\pi e^2 / \epsilon k) $
 are  the Fourier transforms of the Coulomb interactions between electrons
  in the same and different layers respectively.
  All distances (and inverse wave vectors) are in units of the
  magnetic length {\it l}. 
The $ \beta m_{z}^{2}$ term represents the so-called capacitance or
 charging energy needed to maintain unequal amounts of charge density
 in the two layers. Recall that the z-component of pseudospin represents
 the difference between the densities in the two layers. The constant
  $\beta $ is given by 
 \beq  \beta \ = \ \big( {\nu \over 8 \pi^2}\big) \int_{0}^{\infty} dk 
  \ k \ (V^{z}(0) - V^{z}(k)) \ exp({-k^2 \over 2}) \label{beta}   \eeq
where   $V^z_k = {1 \over2} (V^A_k - V^E_k)$. 

\section{The field equation and its solutions}

The nonlinear partial differential equations that have to be obeyed
by all slowly varying spin textures, whether they be single
merons or multi merons is
obtained by extremising this effective action $I[\vec m]$ with respect to 
 $ {\vec m}({\vec r })$. In doing this extremisation one has to respect
 the condition that ${\vec m}$ is a unit vector. We handle this by 
 writing ${\vec m}$ in terms of independent fields $m_z$ and $\alpha$
such that
\beq {\bf m}= \big( \sqrt{1-m_{z}^{2}} \   cos\alpha, \ \ \sqrt{1-m_{z}^{2}}
 \  sin\alpha, \ \ m_{z}) \eeq
 As one can see  $\alpha=tan^{-1}(\frac{m_{y}}{m_{x}})$   
 is the azimuthal angle  of the vector $\vec m$ in the x-y plane.
Upon rewriting $I[\vec m]$ in terms of the independent fields
 $ m_{z}({\vec r })$ and $\alpha(\vec r)$, and minimising with respect to them
 we get the field equations
 
\beq
2\beta m_{z}-\rho_{A}\nabla^{2}m_{z}-\rho_{E} m_{z} \bigg( \frac{(\nabla
m_{z})^{2}}{(1-m_{z}^{2})^{2}}+\frac{m_{z} \nabla^{2}m_{z}}{1-m_{z}^{2}}+
\nabla^{2}\alpha \bigg) \ \\
 + \  \frac{e^{2}d^{2}}{16\pi^{2}\epsilon}\int d^{2}r\int
d^{2} r'(\frac{\nabla^{2}m_{z}({\bf r'})}{|{\bf
r}-{\bf r'}|}) \ =  \ 0 \label{mz} \eeq
and \beq {\vec \nabla} .\big[ (1-m_{z}^{2}){\vec \nabla} \alpha \big]=0 
\label{alpha} \eeq

We will numerically solve these coupled equations for the
variables $m_z$ and $\alpha$ .  In the case of the Skyrmion and bi-meron
solutions people have found it
 helpful to begin with the Isotropic Nonlinear Sigma
model (NLSM) limit of the effective action and field equations,
whose analytical solutions are exactly known. These exact
solutions offer a starting point for setting up  trial
functions 
 for numerically solving the full equations. In the
case of the meron however, there is no exact single meron
solution to the NLSM. To see this let us briefly consider the
NLSM limit.

\subsection{ Absence of single-merons in the NLSM limit}

The SU(2) invariant NLSM limit is obtained when the layer
separation vanishes ($d/l \rightarrow 0$).
At the level of the microcopic Hamiltonian one can see that it is 
rotationally (SU(2) ) invariant in pseudospin in the limit of zero separation.
The potential
energy in (\ref{HI}) and (\ref{int}) is layer independent when d
= 0, and in the LLL approximation the kinetic term (\ref{HK}) involving
the magnetic field is just a constant . This symmetry is also reflected
 in the effective action $I[{\vec m}]$.
  With$d$= 0, we see from their definitions that $\rho_A =
\rho_E $ while $\beta = 0 = C[\vec m]$.  Then the effective
action in (\ref{Eff}) reduces to just the 
 NLSM. The equation for $m_z$ in turn reduces to
\beq   \frac{m_{z}(\nabla
m_{z})^{2}}{(1-m_{z}^{2})^{2}}+\frac{\nabla^{2}m_{z}}{1-m_{z}^{2}}+
m_{z}\nabla^{2}\alpha  = 0 \label{NLSM} \eeq
while \ref{alpha} remains the same. The NLSM has been fully solved 
 \cite{Raj} and  its 
 solutions are most conveniently described in terms of
 the complex field   w(z) defined by
 \beq w(z) \equiv {m_x + im_y \over (1 - m_z)} \eeq
 where z = x+iy. Any analytic function w(z) will be a solution of the NLSM. In
 particular, the single Skyrmion solution is $w(z) = {z \over \lambda}$,
 which corresponds in polar coordinates $(r, \theta)$ to the 
 circularly symmetrical function with
 \beq m_z = {r^2 - \lambda^2 \over r^2 + \lambda^2} \ \ \ \ 
  and \ \ \ \ \ \ \alpha = \theta \label{Skyr} \eeq
It can be checked that (\ref{Skyr}) satisfies  equation (\ref{NLSM}) 

Now ,    isolated meron or anti-meron solutions
 may be defined  by the boundary conditions 
\beqarr m_z(r=0) \  &=&  \ \pm 1 \nonumber \\
   and  \  \ \ \ \ m_z  \ = \  0  \ ; \ m_x +im_y  \ &=&  \ exp(\pm i\theta),
     \ \ \ \  as  \ r  \  \rightarrow  \ \infty \label{bc} \eeqarr
 Solutions with such boundary conditions may or may not exist for a given
 equation. In fact we will argue that the basic NLSM  has no meron solutions.
 But the NLSM does have 
 a Skyrmion solution of the form (\ref{Skyr}).  
The notion of a meron stems from the {\it portion} of the 
Skyrmion function (\ref{Skyr}) 
 from r = 0 to $r = \lambda$. In this portion
 the circle of radius $\lambda$ in the z-plane is mapped onto the
 unit circle in the w-plane. The spin begins at 
 r = 0 with $m_{z} = -1$ starts tilting towards the x-y plane, and by 
 $r = \lambda$ is lying entirely on the x-y plane, pointing radially outward.
The topological number density {\it when integrated over this portion},
 is one-half. Sometimes this portion is called a meron, i.e. half a Skyrmion.
  However,
  it must be emphasized that this object 
corresponds only to  a piece of the full Skyrmion function and is not a
solution to the field equations in the full x-y plane. In contrast 
an "isolated single meron solution"  to the 
field equation, while qualitatively similar,
 should span the whole x-y plane
 and not just a portion of it. That is, it should 
 satisfy the boundary condition (\ref{bc}) at infinity and not at some finite 
 $r = \lambda$.
Such a solution does not exist. The easiest way to understand this is to 
note that there exists no analytical function
w(z) which maps the z-plane into a unit circle in the w-plane.  
Equivalently, if one tried to solve the differential equation 
\ref{NLSM} starting with $m_z = \ -1$ at r=0 and {\it any} positive slope, one
will always end up with a Skyrmion, i.e. with $m_z = +1$ at $r = \infty$.
Note that this is  not directly related to the fact that an isolated meron
,if it existed, would have a logarithmically divergent energy. 
 Solutions can exist  to  diferential equations with some
boundary conditions, even if they have infinite value for the energy 
functional. But in the case of merons, no such solution exists for the NLSM.

\subsection{Meron solutions when $d \neq 0$}

When the layer separation $d$ is non-zero, then the full effective
action (\ref {Eff}) comes into play and we have no longer just the NLSM. The
arguments of the preceding subsection do not hold and now meron
solutions can exist.  See Affleck \cite{Aff} for arguments show
that in fact they will exist when $\beta $ in eq (\ref{mz}) is
non-zero. 

We will look for spherically symmetric solutions with $\alpha (r, \theta)
 \ = \ \theta$ which will satisfy eq(\ref{alpha}) provided $m_z $ depends 
 only on the radius r . Then eq (\ref{mz}) requires that $m_z(r) $ satisfy
\vfill \eject

\begin{eqnarray}
2\beta m_{z}-\rho_{A}(\partial_{r}^{2}m_{z}+\frac{1}{r}\partial_{r}m_{z})
-\rho_{E}\bigg( \ \frac{m_{z}(\partial_{r}m_{z})^{2}}{(1-m_{z}^{2})^{2}}+\frac{m_{z}^
{2}(\partial_{r}^{2}m_{z}+\frac{1}{r}\partial_{r}m_{z})}{1-m_{z}^{2}}+\frac
{m_{z}}{r^{2}} \ \bigg) \nonumber \\
 + \frac{e^{2}d^{2}}{16\pi^{2}\epsilon} \int
d^{2} r'\bigg(\frac{\nabla^{2}m_{z}({\vec r'})}{|{\vec
r}-{\vec r'}|}\bigg)=0 \label{mzeq} \end{eqnarray}  
 
 The parameters  $\rho_A, \  \rho_E$ and $\beta$ in the
equation  are calculated for  each value of the layer separation
$d$ from their definitions (\ref{rho} and \ref{beta}).

  As it stands (\ref{mzeq}) is a nonlinear integro-differential
equation. It is the nonlocal integral in the last term which
represents the most difficult part of this equation.  To start with let
 us neglect this term .  We are
then left with a nonlinear ordinary second order differential equation
given by 
\beq 2\beta
m_{z}-\rho_{A}(\partial_{r}^{2}m_{z}+\frac{1}{r}\partial_{r}m_{z})
-\rho_{E}\bigg(  \ \frac{m_{z}(\partial_{r}m_{z})^{2}}{(1-m_{z}^{2})^{2}}+\frac{m_{z}^
{2}(\partial_{r}^{2}m_{z}+\frac{1}{r}\partial_{r}m_{z})}{1-m_{z}^{2}}+\frac
{m_{z}}{r^{2}}  \ \bigg) \label {mzeq2} \eeq

 We will first solve this equation and then later on return to  the
  full equation 
 (\ref{mzeq}) including the last integral term.  To solve (\ref{mzeq2})
  numerically we merely start with a  boundary
value and slope for $m_z$  at r = 0 , obtain
$\partial_{r}^{2}m_{z}$ using the equation (\ref{mzeq}), use
this to get the value and slope at the next point in r , and
thus proceed towards $r \rightarrow \infty$.  One can see that
the aymptotic behaviour of $m_z(r)$ must be
\beqarr \ \  m_z \ &=& \ 1 - ar^2 + O(r^3)  \ \ \ \ \, as  \ \ r \rightarrow 0 
\nonumber \\
and  \ \ \ \ \ \ \ \ \  m_z  \ &\sim& \ exp(-kr) \ \ \ \ \ \ \ \  
as  \ \ \ r \rightarrow \infty \label{asymp} \eeqarr
where $k \equiv  ( \sqrt{\frac{2\beta}{\rho_{A}}} ) $ and  $a$ is a constant
 for a given layer separation d. The absence of  an $O(r)$ term in 
 $m_z(r \simeq 0)$ 
 follows from the requirement that our solution  not have a cusp at r = 0.
One may be concerned that the last term (proportional to $\rho_E$) in 
eq (\ref{mzeq2}) appears to have pieces with an $r^{-2}$ singularity at r = 0. 
It must
be remembered that $(1 - m^2)^{-1}$  also behaves as $r^{-2}$.  However 
it can be checked that the requirement (\ref{asymp}) ensures that 
the $r^{-2}$ singularities in the different pieces
 cancel  one another, regardless of the value of the constant $a$ .
  This constant  $a$ is adjusted 
 so that $m_z$ exponentially falls to zero as $r \rightarrow
    \infty$ as required.

\section{Results and Discussion}

 The resulting solutions of eq (\ref{mzeq2}) for $m_z(r)$ are
  plotted in figure 1
for a range of 6 different values of $d$ = 0.24 , 0.50 , 0.625 ,
0.78 , 1.00 , 1.20 and 1.50 respectively in units of the
magnetic length {\it l} . Each solution falls monotonically to zero
as $r$ increases with an exponential fall as $r \rightarrow \infty$
as required by(\ref{asymp}). Comparing solutions for different
values of the interlayer distance $d$, we see
 that the width (mean radius)
of the solution monotonically decreases with $d$, the outermost
curve shown in fig.1 corresponding to $d = 0.24$ and the inner
most to $d = 1.50$. The physics behind this is the minimisation
of the capacitance energy $\beta m_z^2$ , which would tend to
bring $m_z$ down to zero as quickly as possible, in contrast to  the
gradient (stiffness) terms would tend to reduce 
$m_z$ as slowly as possible from its starting value of unity at
r = 0. As $d$ increases so does $\beta$  , while $\rho_E$ falls with $d$  and
$\rho_A$ is independent of $d$ (see equations (\ref{beta}) and
\ref{beta}).  Therefore the relative importance of the capacitance
energy to the stiffness terms increases with $d$, and hence the solutions
fall to zero faster.

	In fig.2 we plot the meron core radius $R_{mc}$, which
we define following the literature  \cite{Yang} as the radius at
which $m_z$ equals 0.1. As noted above this radius falls
monotonically with $d$. In fig.2 we show our values of $R_{mc}$
for 8 different values of $d$, while for comparison the results
of Yang and MacDonald (extracted from fig.2 of ref.
\cite{Yang}) are shown in the form of a curve. They consider
their results to be reliable in the region $ 0.5 \leq $d$ \leq
1.2 $. In this range our values of $R_{mc}$ are of the same order
of magnitude as theirs but about 30 to 40 percent smaller . The
qualitative behavior as a function of $d$ is very similar.

	The energy of the meron, i.e. the value of the effective action 
for the meron solution will diverge logarithmically if evaluated over 
the entire plane
because of its nonvanishing azimuthal gradient. However the notion of a 
"meron core energy" applies to the energy of the meron function evaluated 
upto $r  \ = \  R_{mc}$.  \ In fig.3 we plot $E_{mc}$ as a function of $d$.

	Since single merons have , strictly speaking, divergent energy,
physical finite energy excitations involving merons must come in the
form of meron-antimeron pairs, which do have finite energy .
 Using $E_{mc}$ one can  make a phenomenological estimate
 of the energy $E_{MP}$ of such meron pairs as pointed out in
   references \cite{Moon} 
 and \cite{Yang}. Following them we can approximately write the
  energy of a pair separated by a distance R in the form
 \beq E_{MP} \ = \ 2E_{mc} + 2 \pi \rho_{E} ln(R / R_{mc}) + 
 {e^2 \over 4 \epsilon R} \label{MP}\eeq
 The optimal value $R^*$ of
 the meron separation can be obtained by minimising this energy with respect
 to R to get $R^* = {e^2 \over (8\pi \epsilon \rho_E)} $.  Substituting this
 value of $R^*$ into the above equation we get 
 \beq E_{MP} \ = \ 2E_{mc} + 2 \pi \rho_{E} \bigg( 1 \ + \ 
 ln({e^2 \over 8 \epsilon \rho_{E} R_{mc}}) \bigg) \eeq	
Table 1 displays the values of meron core radius, core energy,
the merom pair separation and energy as obtained by us, for
different values of $d$. These may be compared to values
obtained by others using microscopic Hartree Fock methods. We see that for
the lower values od $d$ , such as 0.24 and 0.5, the meron separation
$R^*$ is either smaller or of the same order as the core radius $R_{mc}$.
The approximation of two well separated merons is clearly not
good at these values of $d$. Equations such as (\ref{MP}) should be used
only at layer separation $d$ beyond 0.6. This is in agreement with the
findings of Yang and MacDonald \cite {Yang}.

	Finally we include the non-local integral (the last
term) in the original field equation (\ref{mzeq}) which we had
neglected so far. It arose in this equation as the functional
derivative $ {\delta C[{\vec m}] \over \delta m_z } $ of the
term $C[\vec m]$ in the Effective action (\ref{Eff}). The
presence of this term not only makes the equation more difficult
to solve , but also alters to some extent the nature of the
solution. Notice that this term behaves for large $r$ as ${C
\over r}$ where $C \propto \int d^2 r' \nabla^{2}m_{z}({\vec r'})$.
This ${1 \over r}$ behavior will not permit an exponential fall
off of $m_z$  as had happened in the solutions in the absence of
this term , shown in figure 1. Instead the solution can be expected
to fall to zero asymptotically only as some inverse power of
$r$. Further, the value of the constant C for most values of $d$
of interest is positive. For large $r$ the solution for $m_z$,
may therefore approach zero from below. 

These expectations are supported by the numerical solution of
the integro-differential equation (\ref{mzeq}).  We solve it by
the following iterative procedure.  Start with the solution
$m_z^{0}$ of eq(\ref{mzeq2}) , where the above-mentioned
integral has been altogether neglected. Evaluate the integral by
inserting $m_z^{0}$ in its integrand. Multiply the resulting
integral by a prefactor $\alpha = 0.1$  , add it to
eq.(\ref{mzeq2}), and solve the equation again to obtain
$m_z^{\alpha}$. Solve the equation again after inserting this
$m_z^{\alpha}$ into the integral, call the solution
$m_z^{2\alpha}$, and repeat this procedure ten times until you
get $m_z^{10\alpha}$. Since  $ 10 \alpha = 1$ this should yield
a good approximation  to the full equation (\ref{mzeq}). One
could have used this iterative procedure in smaller steps , but
we find the step size of $\alpha = 0.1$ is small enough for the
procedure to converge  smoothly .

In figure 4 we compare the solution so obtained for the full
equation (\ref{mzeq}) with the "unperturbed" solution (of the
simpler eq (\ref{mzeq2}) discussed earlier), for two values of
$d$ equal to 0.625 and 1.20.  One can see that the full solution
reaches zero, and then after turning negative, returns to zero
asymptotically but more slowly than the unperturbed solution.
In other words the pseudospin, which starts out pointing along
 the z-axis at r=0,
gradually falls towards the x-y plane, lies entirely on the x-y plane at
some finite radius, then starts bending towards the negative z-axis
, then reverses its behavior to finally asymptotically come back
to the x-y plane at infinity. We see no objection to such a meron solution.
But the notion of a core may be less meaningful now.
One could formally evaluate the "meron core radius $R_mc$ " of
this new solution defined once again as the radius at which
$m_z = 0.1$, and a corresponding core energy. But with the
solution continuing to be substantially nonzero at radii quite a
bit larger than such an $R_mc$ , the concept of a core is less
significant for such slowly falling solutions.

\vspace{1.0 cm}

\underbar{{\bf Acknowledgement}}

R.R acknowledges many valuable discussions with Dr Shivaji
Sondhi on Hall systems. We also thank Dr Subir Sarkar for 
his unstinting help in doing the numerical
work here. The work of S.G. is supported by a
CSIR award no.9/263(225)/94-EMR-I .dt.2.9.1994.
\newpage

\noindent Table 1: Meron core radius ($R_{mc}$), meron-antimron separation
 ($R^{*}$),meron core energy( $E_{mc}$ ) and meron pair energy ($E_{MP}$) as
  a function of layer separation.
The unit of energy is $\frac{e^{2}}{\epsilon l}$ and the unit of length
is ${\it l}$

\begin{center}
\begin{tabular}{|c|c|c|c|c|}
\hline
d &  $R_{mc}$ & $R^{*}$ & $E_{mc}$ & 
$E_{MP}$ \\ 
\hline
.2400 & 5.07276 & 2.3133 & 0.145487 & 0.314187 \\
\hline
.5000 & 3.07692 & 3.52110 & 0.105005  & 0.290580\\
\hline
.6250 &2.70270 & 3.9868 & 0.101588 & 0.290250 \\
\hline
.7810 & 1.99584 & 4.88985 & 0.084911 & 0.266760 \\
\hline
1.000 & 1.41372 & 6.42600  & 0.067163 & 0.232140\\
\hline
1.200 & 1.14761 & 8.15756 & 0.057474 & 0.205750 \\
\hline
1.500 & 0.873181 & 11.3682 & 0.042215 & 0.163039 \\
\hline 
\end{tabular}\\
\end{center}

\begin{figure}
\label{fig1}
\caption{The solution $m_{z}(r)$ of equation(\ref{mzeq2}).
 The  six curves correspond  to layer separation $d$ equal to  0.24 , 0.50 ,
  0.625 , 0.78 , 1.00 , 1.20 and 1.50 respectively in units of the
magnetic length {\it l}. The width of the curves decreases monotonically as
as $d$ increases, with the outermost curve for $d$ = 0.24 and the innermost to
 $d$ = 1.50}
\end{figure}
\begin{figure}
\label{fig2}
\caption{Meron core radius $R_{mc}$ for 8 different values of the
 layer separation $d$ , all in units of ${\it l}$. The solid curve gives
  the results obtained in ref.10} 
\end{figure}
\begin{figure}
\label{fig3}
\caption{Meron core energy $E_{mc}$ as a function of layer separation}
\end{figure}
\begin{figure}
\label{fig4}
\caption{$m_{z}(r)$ versus r. The continuous lines give the solution
 of eq(\ref {mzeq}) which includes the 
integral term  and, for comparison, the dotted curves depict
the solution of (\ref{mzeq2}) which does not include this term.  Figure (a)
corresponds to d = 0.625 and figure (b) to d = 1.2}
\end{figure}

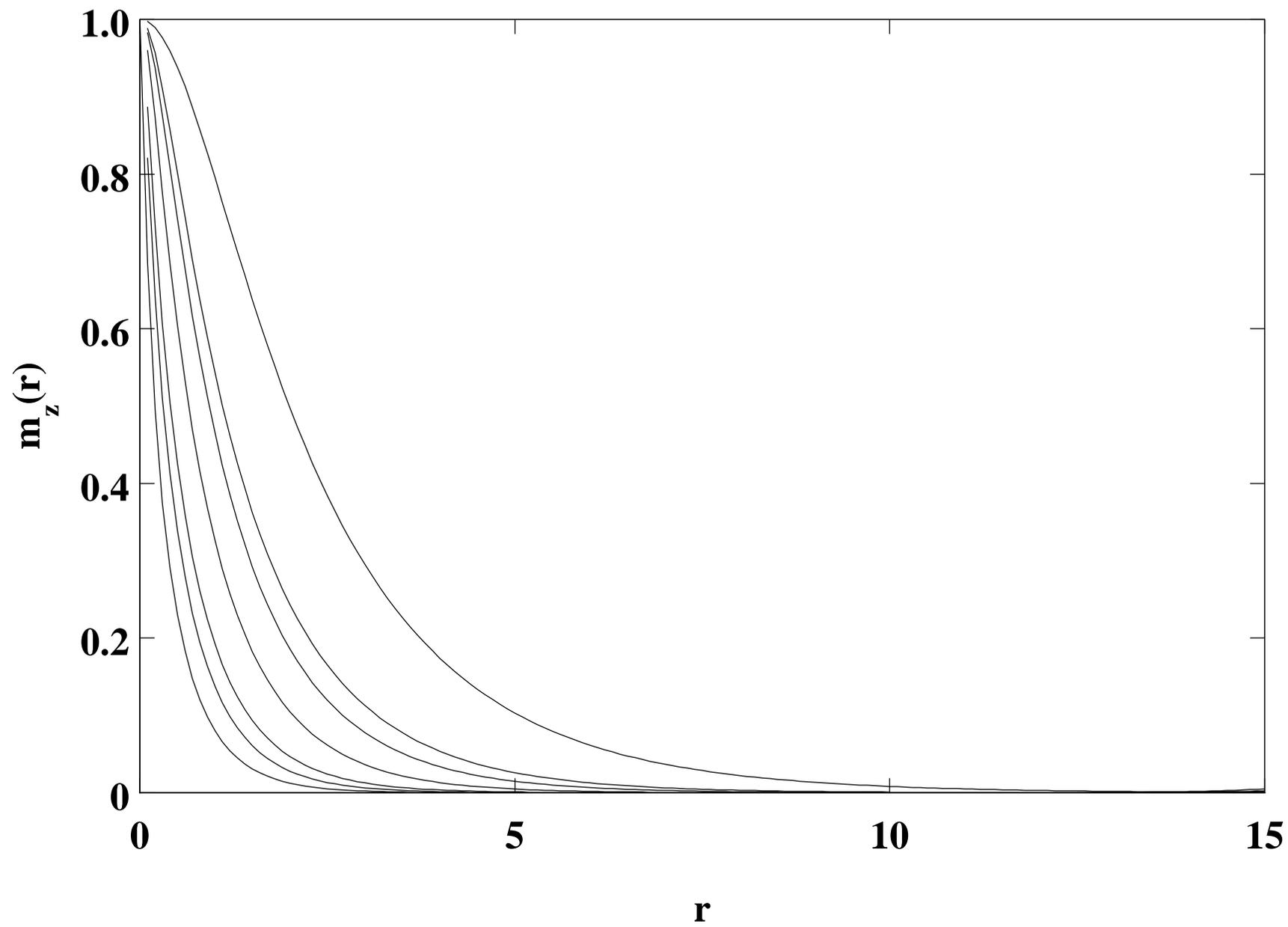
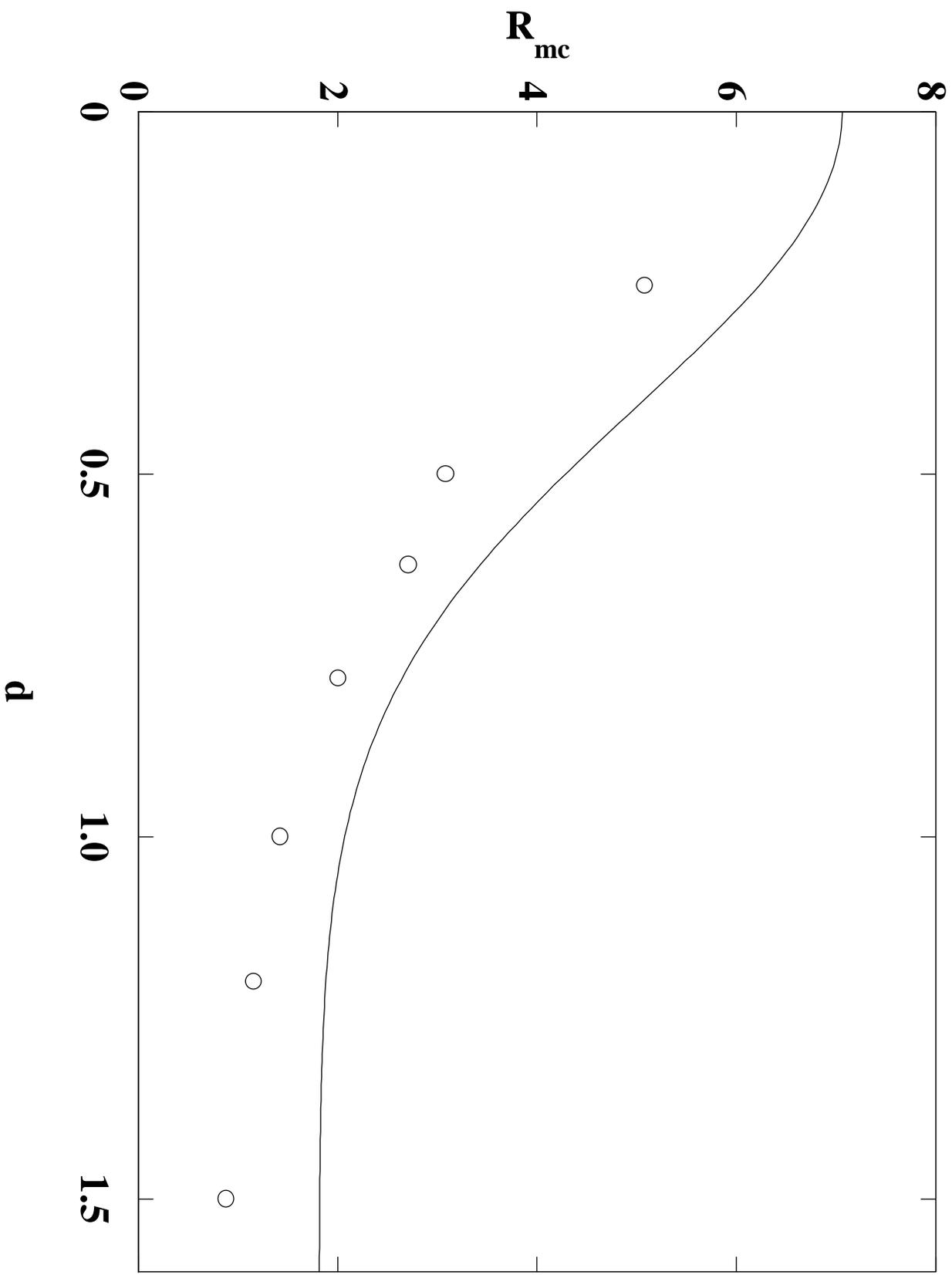
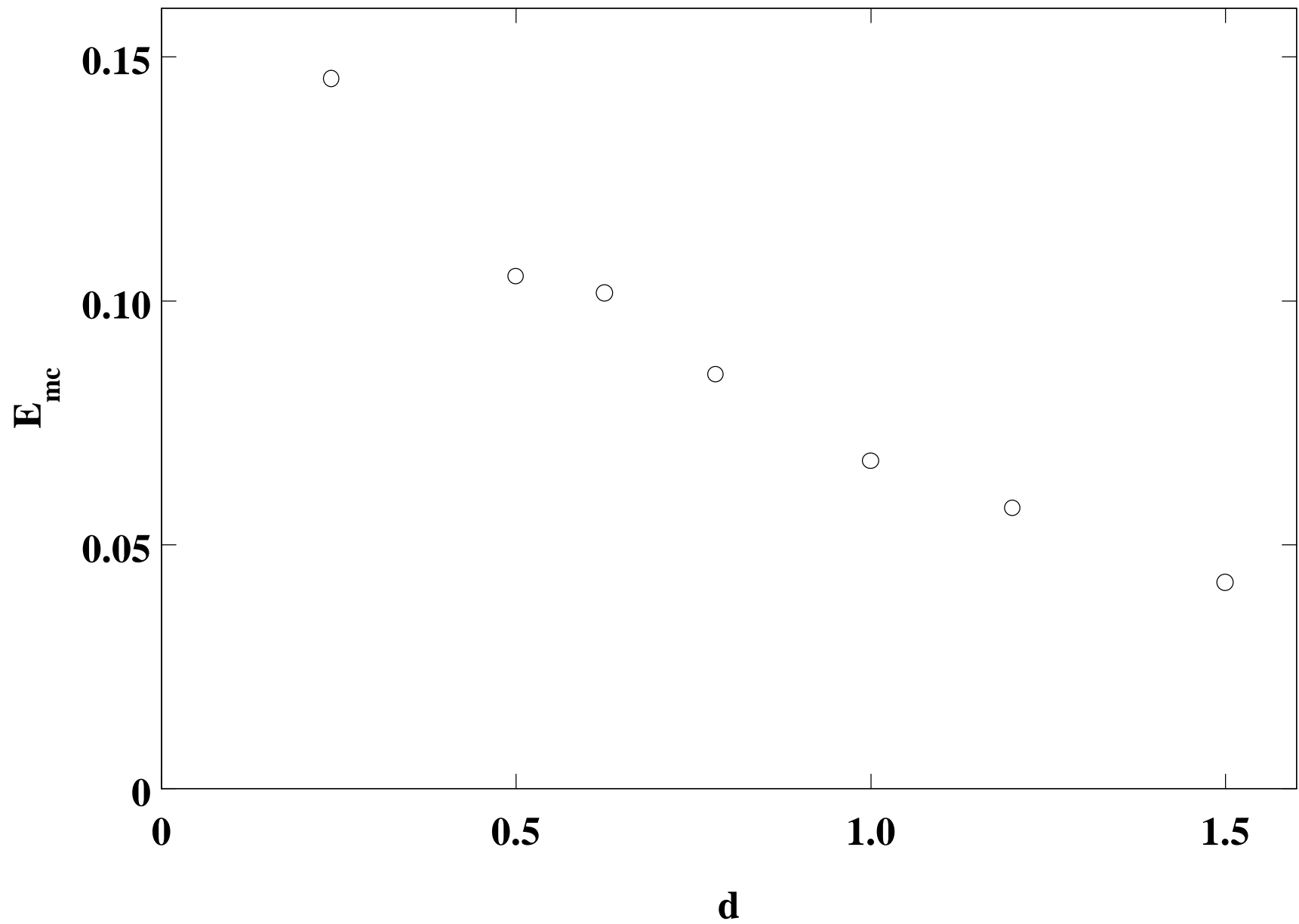
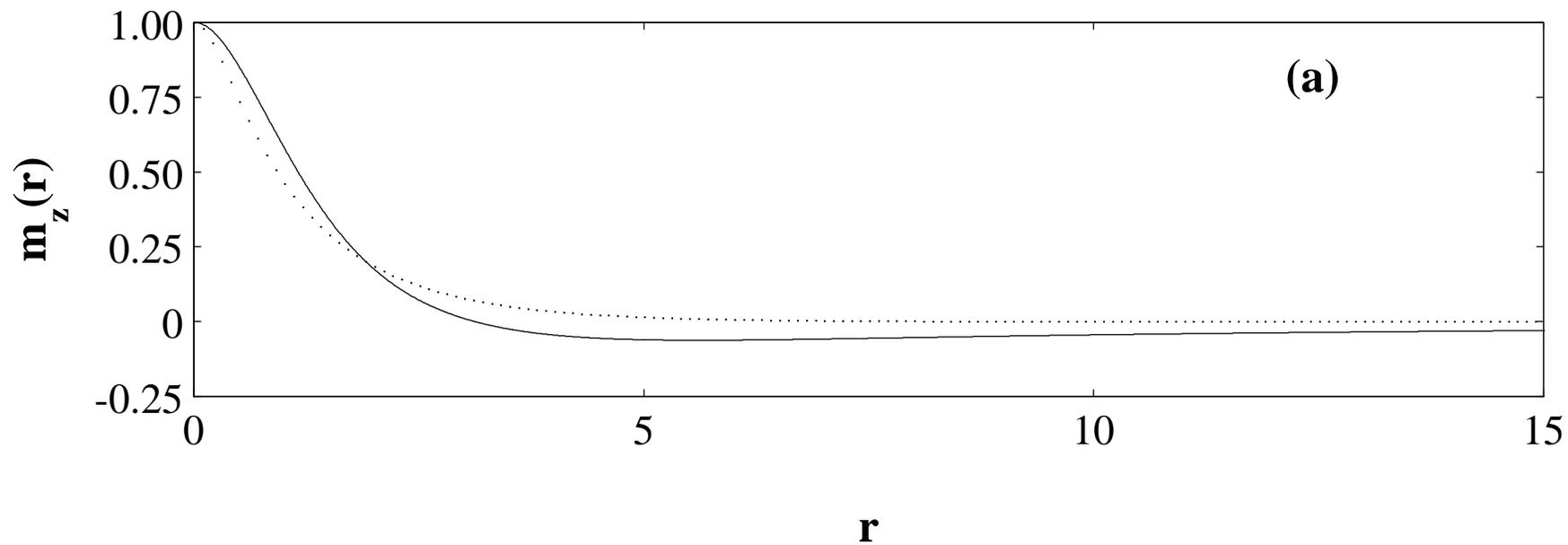
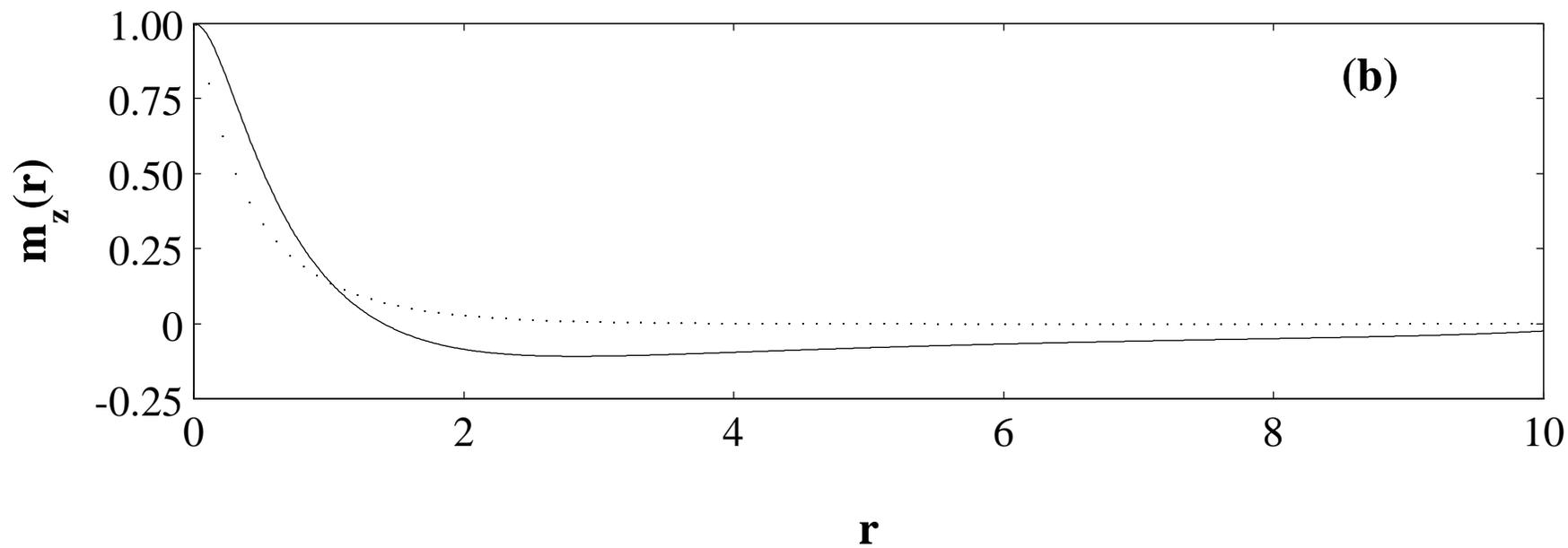
\end{document}